\documentclass[fleqn,10pt]{wlscirep}
\usepackage[utf8]{inputenc}
\usepackage[T1]{fontenc}

\usepackage{graphicx}  
\usepackage{url}           
\usepackage{dcolumn}       
\usepackage{bm}            
\usepackage{amssymb}       
\usepackage{mathtools}     
\DeclarePairedDelimiter\ket{\lvert}{\rangle}
\usepackage{amsmath}       
\usepackage{pgf}          
\usepackage{tikz,pgfplots}
\usepackage[super, sort&compress]{natbib}
\usepackage{chngcntr}
\usetikzlibrary{shapes,arrows,calc}
\usetikzlibrary{arrows, decorations.markings}
\usetikzlibrary{decorations.pathreplacing,calligraphy}

\setcitestyle{citesep={,}}

\definecolor{green}{RGB}{34,139,34}

\begin{document}

\title{Spectral quantum algorithm for passive scalar transport in shear flows}

\author[1,*,+]{Philipp Pfeffer}

\author[2,3,+]{Peter Brearley}

\author[2]{Sylvain Laizet}

\author[1]{J\"org Schumacher}
\affil[1]{Institute of Thermodynamics and Fluid Mechanics, Technische Universit\"at Ilmenau, Postfach 100565, D-98684 Ilmenau, Germany}
\affil[2]{Department of Aeronautics, Imperial College London, London SW7 2AZ, UK}
\affil[3]{Department of Mechanical and Aerospace Engineering, University of Manchester, Manchester M13 9PL, UK}

\affil[*]{philipp.pfeffer@tu-ilmenau.de}

\affil[+]{These authors contributed equally to this work}

\begin{abstract}
    The mixing of scalar substances in fluid flows by stirring and diffusion is ubiquitous in natural flows, chemical engineering, and microfluidic drug delivery. Here, we present a spectral quantum algorithm for scalar mixing by solving the advection-diffusion equation in a quantum computational fluid dynamics framework. The exact gate decompositions of the advection and diffusion operators in spectral space are derived. For all but the simplest one-dimensional flows, these operators do not commute. Therefore, we use operator splitting to construct quantum circuits capable of simulating arbitrary polynomial velocity profiles in multiple dimensions, such as the Blasius profile of a laminar boundary layer. Periodic, Neumann, and Dirichlet boundary conditions can be imposed with the appropriate quantum spectral transform. We evaluate the approach in statevector simulations of a Couette flow, plane Poiseuille flow, and a polynomial Blasius profile approximation. For an advection-diffusion problem in one dimension, we compare the time evolution of an ideal quantum simulation with those of real quantum computers with superconducting and trapped-ion qubits. The required number of two-qubit gates grows with the logarithm of the number of grid points raised to one higher power than the order of the polynomial velocity profile. 
\end{abstract}

\maketitle

\section{Introduction}

The theoretical foundations for quantum computing were laid in the early 1980s by Benioff \cite{Benioff1980} and Feynman,\cite{Feynman1982} but it was not until the 1990s that the field captured widespread attention with Shor's algorithm for integer factorization.\cite{Shor1994} The emergence of quantum hardware capable of proof-of-concept computations \cite{Arute2019} has further added to the momentum, yet the number of industries that are set to benefit from quantum computing remains disappointingly few.\cite{Au-Yeung2024} Most modern supercomputing resources are dedicated to the numerical simulation of partial differential equations (PDEs), where finding analytical solutions is often either impractical or impossible. Developing quantum algorithms for these applications offers a practical way to advance the broader benefits of quantum computing technology. Quantum algorithms for simulating the unitary dynamics governed by the Schr\"odinger equation $i\, d\vec{\phi}/dt = H\vec{\phi}$ with Hermitian $H=H^\dagger$, collectively known as Hamiltonian simulation algorithms, are well-established,\cite{Low2019} with applications in quantum chemistry, materials, and particle physics. The focus in this work therefore shifts to the dynamics governing non-unitary evolutions where $H\ne H^\dagger$, appearing broadly across science and engineering in systems that exhibit dissipation or instability.

A prominent example of a non-unitary PDE is the advection-diffusion equation for the mixing of a substance in a fluid flow,\cite{Villermaux2019} which is given by
\begin{equation}
    \frac{\partial \phi}{\partial t} + \vec{u} \cdot \nabla \phi = D \nabla^2 \phi.
    \label{eq:advection_diffusion}
\end{equation}
The equation describes how a passive scalar field $\phi(\vec{x},t)$ is advected by an incompressible (divergence-free) velocity field $\vec{u}(\vec{x},t)$ subjected to diffusion characterized by a constant diffusivity $D$. Its coupling with the incompressible Navier–Stokes equations, given by
\begin{equation}
    \frac{\partial \vec{u}}{\partial t} + \vec{u} \cdot \nabla \vec{u} = -\frac{1}{\rho} \nabla p + \nu \nabla^2 \vec{u}\,,
    \label{eq:momentum_equation}
\end{equation}
with the condition $\nabla \cdot \vec{u} = 0$ and appropriate boundary conditions, is essential to fluid simulations used in the aeronautical, automotive, chemical and renewable energy industries. In Eq.~\eqref{eq:momentum_equation}, the introduced quantities are density $\rho$, pressure $p$, and kinematic viscosity $\nu$. In low Reynolds number flows, where $\|\nu \nabla^2\vec{u}\| \gg \|\vec{u}\cdot \nabla\vec{u}\|$ and thus viscous forces dominate, the velocity field is laminar and can be obtained in analytical form without the need to solve Eq.~\eqref{eq:momentum_equation} numerically. These flows are typical problems in microfluidics\cite{Bruus2007} and lubrication systems\cite{Elrod1979} where the characteristic length and velocity scales are small. Although the velocity field is known, predicting the evolution of scalar fields such as concentration or temperature by Eq.~\eqref{eq:advection_diffusion} remains non-trivial, particularly for $\nu\gg D$.

By discretizing the spatial derivatives of Eq.~\eqref{eq:advection_diffusion}, a system of ordinary differential equations (ODEs) $d\vec{\phi}/dt = M\vec{\phi}$ for the discrete right-hand-side operator $M$ is produced that can be targeted by general linear ODE solvers.\cite{An2023, Jin2023, Fang2023, Berry2024} The central objective for a quantum algorithm is to efficiently transform a non-unitary ODE into an equivalent unitary form governed by the Schr\"odinger equation. Several pathways have been proposed in recent years, e.g., the linear combination of Hamiltonian simulation\cite{An2023} or Schr\"odingerization techniques.\cite{Jin2023} An insightful perspective comes from the decomposition of $M$ into anti-Hermitian (unitary) and Hermitian (non-unitary) components. The simulation of the Hermitian components is made possible by embedding the evolution operator into a larger and overall unitary system, which is known as block-encoding.\cite{Fang2023} The probability to extract this embedded output through measurement becomes an important aspect and can be improved by amplitude amplification techniques,\cite{Brassard2002} such as the repeated application of uniform singular value amplifications.\cite{Gilyen2019} However, these techniques increases the circuit depth, for instance to a quadratic dependence on the evolution time.\cite{Fang2023} Another direction is based on a truncated Dyson series of the ODE, which provides the solution of the ordinary differential equation in the form of a time-ordered integral,\cite{Berry2024} and may be encoded in a linear system of equations and solved with a quantum linear systems algorithm (QLSA).\cite{Harrow2009}

To overcome the shortfalls of general quantum linear ODE solvers, algorithms specifically targeting the advection-diffusion equation,\cite{Budinski2021, Demirdjian2022, Bharadwaj2023, Jaksch2023, Ingelmann2024,Over2025b,Over2025a,Bharadwaj2025} as well as its constituent advection equation \cite{Brearley2024, Sato2024, Au-Yeung2025} and heat equation \cite{Linden2022} have been developed in recent years. This has occurred in the context of the lattice Boltzmann method,\cite{Budinski2021} variational quantum algorithms,\cite{Demirdjian2022, Ingelmann2024, Jaksch2023} implicit time marching with a QLSA,\cite{Bharadwaj2023, Ingelmann2024,Bharadwaj2025} explicit time marching by block encoding,\cite{Over2025b} Hamiltonian simulation,\cite{Brearley2024, Sato2024} and smoothed-particle hydrodynamics.\cite{Au-Yeung2025} However, these algorithms typically require deep or many circuits, which presents a significant challenge to demonstrate near-term quantum advantage.

In this study, we present a quantum spectral method for solving the advection-diffusion equation, motivated by the efficient and well-known implementation of the quantum Fourier transform (QFT), and the exponentially improved accuracy compared to other methods. Fourier methods have been applied previously to solve partial differential equations, such as linear \cite{Wright2024,Lubasch2025} and non-linear \cite{Koecher2025} wave equations or incompressible hydrodynamic Schr\"odinger equations.\cite{Meng2023} We present simple quantum circuits for simulating the constituent processes of advection and diffusion in Fourier space, each requiring $O(\log^2 N)$ gates to implement in one dimension, where $N$ is the number of grid points in the simulation. When both advection and diffusion can be diagonalized by the same spectral transform, the sequential implementation of the individual algorithms in spectral space corresponds to a simulation of the combined advection-diffusion dynamics, achieving the full exponential solution convergence of spectral methods. We extend this to wall-bounded laminar shear flows under an $x$ velocity $u(y)$, with $x$ and $y$ being the streamwise and wall-normal directions, respectively. Under these conditions, advection and diffusion diagonalize under different multidimensional spectral operators, so operator splitting, such as described by the Lie-Trotter product formula, is required.  Despite the reduction in accuracy from exponential to polynomial, e.g.\ from the first-order convergence of the Lie-Trotter product formula, the step size remains independent of the computational grid unlike in explicit time-marching methods, thereby retaining the exponential advantage in $N$. The overheads of the required number of operator splitting steps are also small, which we show by achieving the accuracy of a 10\textsuperscript{th}-order finite difference solution with four second-order steps per characteristic time scale, where the simulations are performed on equivalent grids. Therefore, we find this to be a promising strategy for simulating scalar transport processes on near-term quantum computers.   

The following two sections will detail the algorithms for the individual implementation of advection and diffusion. Then, details of their unification by operator splitting will be discussed. Numerical examples of scalar transport in one- and two-dimensional flows are presented in the following section, followed by the concluding remarks in the final section.

\section{Spectral quantum advection step}
The advection equation, $\partial \phi / \partial t + \vec{u} \cdot \nabla \phi = 0$, inherently preserves the $L^2$-norm of the passive scalar $\phi$ under an incompressible velocity field $\vec{u}$, provided that the overall flux across the boundaries vanishes. To see this, consider the time derivative of the squared $L^2$-norm over a domain $\Omega$ as
\begin{align}
    \frac{d}{dt} \int_\Omega \phi^2 \, dV = 2 \int_\Omega \phi\, \frac{\partial \phi}{\partial t} \, dV 
    = -\int_\Omega \vec{u} \cdot \nabla (\phi^2) \, dV
    = - \int_{\Omega} \nabla \cdot (\phi^2 \vec{u}) \, dV + \int_{\Omega} \phi^2 (\nabla \cdot \vec{u}) \, dV.
\end{align}
Applying the divergence theorem and noting that $\nabla \cdot \vec{u} = 0$, this reduces to
\begin{equation}
    \frac{d}{dt} \int_\Omega \phi^2 \, dV = \int_{\partial V} \phi^2 (\vec{u} \cdot \vec{n}) \, dS\,,
\end{equation}
which vanishes for boundary conditions that nullify the flux across the boundary, such as periodic or insulating wall conditions. Therefore, advection is inherently a unitary process. However, when the problem is discretized in space to facilitate computation, discretization errors may result in a non-unitary evolution, e.g.\ for multidirectional velocity fields.\cite{Brearley2024} For a constant one-dimensional velocity, the unitary evolution is retained by central finite difference and spectral methods of discretization.

\subsection{One-dimensional case}
The advection equation under a single, constant velocity component $u$ on a periodic domain is a Hamiltonian system in spectral space. The QFT maps the scalar in physical space to Fourier modes with wavenumbers $k_j$ defined and ordered as
\begin{equation}
    k_j = \frac{2\pi}{L}
    \begin{cases}
        j & \text{when } 0\leq j<\frac{N}{2} \\
        j-N & \text{when } \frac{N}{2}\leq j<N,
    \end{cases}
    \label{eq:wavenumber}
\end{equation}
where $N=2^n$ is the number of grid points, $L$ is the length of the domain, and $n$ is the number of qubits encoding the solution. For a single Fourier mode $\hat{\phi}_j(t)$ with wavenumber $k_j$, the advection equation is $d\hat{\phi}_j/dt = -i u k_j \hat{\phi}_j$ with solution $\hat{\phi}_j(t) = e^{-i u k_j t} \hat{\phi}_j(0)$. A Hamiltonian $H = \text{diag}(u k_j)$ can be constructed to simulate the discrete evolution by $\ket{\hat{\phi}(t)} = e^{-i H t} \ket{\hat{\phi}(0)}$. A quantum circuit for this evolution is derived as follows. 
\begin{figure}
    \centering
    \includegraphics[width=0.8\textwidth]{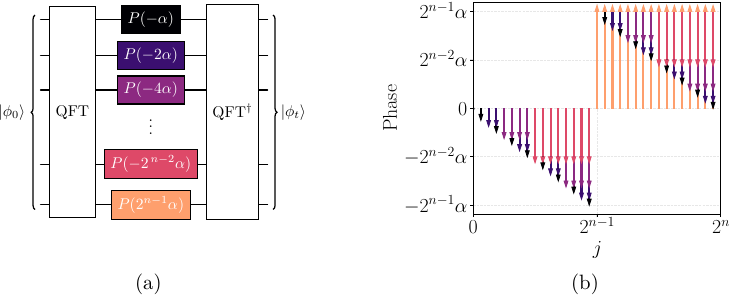}
    \caption{(a) Quantum circuit to implement the advection evolution operator on $n$ qubits for a constant velocity $u$ in a periodic domain by diagonalization under the QFT with $\alpha = 2\pi u t/L$. (b) Illustration of the action of the phase gates by the circuit with a color coding that corresponds to the gates in panel (a).}
    \label{fig:adv_spectral}
\end{figure}
The index $j$ in Eq.~\eqref{eq:wavenumber} has the following binary expansion for $n$ qubits
\begin{equation}
    j = \sum_{r=0}^{n-1} 2^{r}q_r,
    \label{eq:binary_expansion}
\end{equation}
where $q_r\in\{0,1\}$ is the $r$\textsuperscript{th} qubit and $q_0$ is the least-significant qubit appearing at the top of circuit diagrams (little endian). Rewriting $e^{-iuk_jt}$ in terms of the wavenumber definition in Eq.~\eqref{eq:wavenumber} and the binary expansion of $j$ in Eq.~\eqref{eq:binary_expansion} results in 
\begin{equation}
    e^{-iuk_jt} = 
    \begin{cases}
        \displaystyle\exp\left(-i\alpha\sum_{r=0}^{n-1}2^r q_r\right) & \text{when } 0\leq j < \frac{N}{2}, \\
        \displaystyle\exp\left(-i\alpha\left[\sum_{r=0}^{n-1}2^r q_r -2^n \right]\right) & \text{when } \frac{N}{2}\leq j < N,
    \end{cases}
    \label{eq:1d_advection_operator}
\end{equation}
where $\alpha = 2\pi u t/L$ is the number of advective time scales multiplied by $2\pi$. Since the system has been diagonalized in Fourier space, the scalar identity $e^{a+b}=e^ae^b$ can be applied so that the evolution operator can be written as the product
\begin{align}
    e^{-iuk_jt} &= e^{i\alpha2^n q_{n-1}}\prod_{r=0}^{n-1}e^{-i\alpha2^r q_r} = e^{i\alpha2^{n-1} q_{n-1}}\prod_{r=0}^{n-2}e^{-i\alpha2^r q_r},
    \label{Eq:Adv}
\end{align}
and can be implemented by the quantum circuit shown in Fig.~\ref{fig:adv_spectral}a where $P(\theta) = \text{diag}(1, e^{i\theta})$ is the phase gate. Since the $P$ gate is equivalent to the $R_Z(\theta) = \text{diag}(e^{-i\theta/2}, e^{i\theta/2})$ gate up to a global phase, i.e.\ $R_Z(\theta) = e^{-i\theta/2} P(\theta)$, the $P$ gates may be substituted for $R_Z$ gates with no observable impact. Fig.~\ref{fig:adv_spectral}b illustrates how these gates construct the corresponding rotation via Eq.~\eqref{Eq:Adv}. This linear phase gradient with the piecewise shift requires a single gate per qubit in Fourier space, so $n=\log(N)$ gates. The piecewise shift introduced in Eq.~\eqref{eq:wavenumber} is implemented by the positive phase acting on the most-significant qubit in Fig.~\ref{fig:adv_spectral}, as derived in Eq.~\eqref{Eq:Adv}. The circuit for implementing the exact QFT requires $O(n^2)$ gates, while circuits for computing the QFT to accuracy $\epsilon$ requires $O[n\log (n/\epsilon)]$ gates.\cite{Cleve2000} Using the latter implementation results in an overall gate complexity of $O(n\log [n/\epsilon]) = O(\log N [\log\log N+ \log \{1/\epsilon\}])$ gates. This is independent of the advection velocity or simulation time, as these quantities are captured in the phase rotations. The solution converges exponentially in $N$ resulting from the Fourier spectral method, so $N = O(\log\{1/\epsilon\})$. Therefore, the gate complexity can be written entirely in terms of the error as $O(\log [1/\epsilon]\log \log [1/\epsilon])$.

\subsection{Advection in laminar shear flow}

In laminar wall-bounded flows, the velocity in the streamwise $x$ direction $u(y)$ varies as a function of the wall-normal coordinate $y$, with the remaining velocity components $v = w = 0$. For example, a Couette flow consists of two flat parallel plates separated by a distance $L$, where the bottom plate is stationary and the top plate is moving at a velocity of $U$. This type of flow is commonly found in lubrication systems. The moving wall imparts a uniform shear on the fluid with velocity $u(y) = U y/L$. Another paradigmatic case is a Poiseuille flow, where a fluid flows between stationary walls in a pipe or channel driven by a uniform pressure gradient. Here, the velocity profile is parabolic and equal to $u(y) = 4U(y/L)[1-(y/L)^2]$ in the case of a channel with the centerline velocity $U$. In laminar boundary layer flows, the velocity varies from zero at the wall to the free-stream value $U$ over a thin region adjacent to the surface.\cite{Schlichting2016} For a steady, incompressible laminar flow over a flat plate with no pressure gradient, the Blasius solution describes the boundary layer profile.\cite{Schlichting2016} A common polynomial approximation with third-order accuracy is given by $u(y)=2U(y/\delta)-U(y/\delta)^2$ with the boundary layer thickness $\delta$. This parabolic profile satisfies the no-slip condition at the wall and smoothly transitions to the homogeneous free-stream velocity $U$. In the following, a methodology for simulating the advection-diffusion equation in shear flows with polynomial velocity profiles is presented, using the three introduced flows as examples, which are typical of two-dimensional mixing problems.

\begin{figure}[t]
    \centering
    \includegraphics[width=0.99\textwidth]{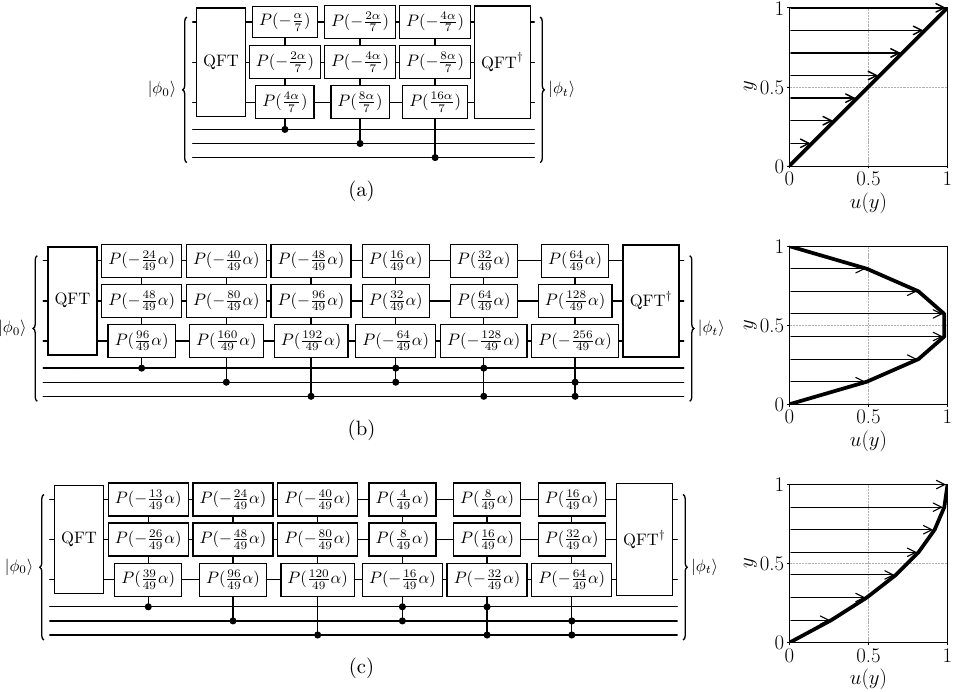}
    \caption{Quantum circuits using $n=3$ qubits per spatial dimension for simulating the advection equation (a) in a Couette flow, (b) a Poiseuille (channel) flow, and (c) a boundary layer described by a third-order approximation to the Blasius solution. The top three qubits correspond to the $x$ direction and the bottom three qubits correspond to the $y$ direction. The phase coefficients for the Couette, Poiseuille and Blasius profiles are calculated from Eqs.~\eqref{eq:couette}, \eqref{eq:poiseuille}, and \eqref{eq:blasius}, respectively. Here, $\alpha=2\pi U t / L$ as the variable velocity profile $u(y)$ is encoded in the prefactors of $\alpha$ in each gate.}
    \label{fig:advection_shear_circuits}
\end{figure}

Since $u(y)$ is the sole non-trivial velocity component, only the qubits corresponding to the $x$ register are required to enter Fourier space. Then, the $y$ register in physical space can be used to control the polynomial velocity profile $u(y)$. The binary expansion of the wavenumber index $j$ in Eq.~\eqref{eq:binary_expansion} can be used to define the wall-normal coordinate $y = (\sum_{r=0}^{n-1}2^rq_r)/(2^n-1)$ as a binary fraction. In all cases, we consider the non-dimensional parameters $L=1$ or $\delta=1$ together with $U=1$ for simplicity. Non-unity values only scale the applied phases, having no impact on the efficiency, stability or accuracy of the simulation.

For a Couette flow where $u(y) = y$, the required phase operations to implement the velocity profile are 
\begin{equation}
    e^{-iu(y)k_jt} = \exp\left(-ik_jt \frac{\sum_{r=0}^{n-1} 2^{r}q_r}{2^{n}-1}\right) = \prod_{r=0}^{n-1} \exp\left( -ik_jt \frac{2^r}{2^{n}-1} q_r \right).
    \label{eq:couette}
\end{equation}
The wavenumber $k_j$ has been left in its unexpanded form to avoid unnecessarily lengthy equations, since following the same procedure in Eq.~\eqref{eq:1d_advection_operator} results in the same phase pattern on the $x$ register as the one-dimensional case in Fig.~\ref{fig:adv_spectral}.  A quantum circuit for solving the advection-diffusion equation of a Couette flow by implementing this strategy on $n=3$ qubits per spatial dimension is shown in Fig.~\ref{fig:advection_shear_circuits}a. Since the velocity field is now variable, we redefine the constant factor $\alpha = 2\pi Ut/L$ as the effects of the variable velocity profile $u(y)$ and $k_j$ are now encoded in the coefficients. The domain is periodic in $x$ with no boundary condition required in $y$ since $v=0$. 
The construction requires $O(\log N)$ control qubit configurations on the $y$ register and $O(\log N)$ phase gates per control configuration on the $x$ register, resulting in an overall gate complexity of $O(\log^2 N)$ that matches the exact QFT implementation.

For a plane Poiseuille (channel) flow, we use $u(y) = 4y(1-y)$. The required phase operations to implement are therefore
\begin{align}
    \exp\left( -ik_jt \frac{4\sum_{r=0}^{n-1} 2^{r}q_r}{2^n-1} \left[1-\frac{\sum_{r=0}^{n-1} 2^{r}q_r}{2^n-1}\right] \right)
    &= \prod_{r=0}^{n-1} \exp\left(-ik_jt \frac{2^{2+r}(2^n-1) - 2^{2+2r}}{(2^{n}-1)^2}q_r\right) \prod_{s>r}^{n-1}\exp\left(ik_jt\frac{2^{3+r+s}}{(2^{n}-1)^2}q_rq_s \right),
    \label{eq:poiseuille}
\end{align}
which is demonstrated in Fig.~\ref{fig:advection_shear_circuits}b. Since $ y^2=(\sum_{r=0}^{n-1} 2^{r}q_r/[2^n-1])^2$ can produce at most $O(n^2)$ terms, the velocity field encoding requires $O(\log^2N)$ control qubit configurations on the $y$ register, resulting in the overall two-qubit gate complexity of $O(\log^3N)$ when accounting for the $O(\log N)$ gates per control configuration on the $x$ register. 

The velocity profile of a laminar boundary layer is obtained by solving the Blasius equation. This has no exact polynomial solution, but a commonly used Pohlhausen polynomial to third-order accuracy is $u(y) = 2y - y^2$. Therefore, the required phase transformation is
\begin{align} 
    \exp\left( -ik_jt \left[\frac{2\sum_{r=0}^{n-1} 2^{r}q_r}{2^n-1} - \frac{(\sum_{r=0}^{n-1} 2^{r}q_r)^2}{(2^n-1)^2} \right]\right) 
    &= \prod_{r=0}^{n-1} \exp\left(-ik_jt\frac{2^{1+r}(2^n-1)-2^{2r}}{(2^n-1)^2}q_r\right)\prod_{s>r}^{n-1}\exp\left( ik_jt \frac{2^{1+r+s}}{(2^n-1)^2}q_rq_s \right),
    \label{eq:blasius}
\end{align}
where the quantum circuit for solving the advection equation with this velocity profile is shown in Fig.~\ref{fig:advection_shear_circuits}c. Since the velocity field is also a quadratic function, $O(\log^3N)$ gates are required.

The technique of encoding a polynomial velocity profile has been demonstrated using common linear and parabolic velocity profiles, although this technique extends to arbitrary $h$-order polynomials requiring $O(h\log^{h+1} N)$ two-qubit gates to implement.\cite{Barenco1995} The Blasius profile may be simulated to the desired order of accuracy using this general strategy.

\section{Spectral quantum diffusion step}

The heat equation, $\partial\phi/\partial t = D\nabla^2\phi$, describes the diffusion of a passive scalar field in a medium at rest with a constant diffusivity $D$. Unlike the advection equation, diffusive dynamics are inherently non-unitary, which poses a fundamental challenge for simulating diffusion on quantum computers. The requirement for a constant diffusivity is typical for diffusion under modest temperature variations, and causes the Laplacian term to diagonalize under spectral transforms with a uniform grid, enabling the construction of efficient spectral quantum circuits. In this section, we will extend the methodologies presented in the previous section to simulate diffusion in a probabilistic framework.

\subsection{Quantum circuit with spectral accuracy}

For a single spectral mode $\hat{\phi}_j$ with wavenumber $k_j$, the one-dimensional heat equation for periodic boundary conditions simplifies to
\begin{equation}
\frac{d\hat{\phi}_j}{dt} = -Dk_j^2 \hat{\phi}_j,
\label{eq:spectral_heat_equation}
\end{equation}
with the solution $\hat{\phi}_j(t) = \exp(-Dk_j^2 t) \hat{\phi}_j(0)$. By defining the diagonal positive semi-definite $H = \text{diag}(Dk_j^2)$, Eq.~\eqref{eq:spectral_heat_equation} can be written in matrix form $d\hat{\phi}/dt = -H\hat{\phi}$ with the solution $\hat{\phi}(t) = e^{-H t} \hat{\phi}(0)$. In a quantum mechanical framework, this could be implemented by imaginary time evolution algorithms\cite{Motta2020, McArdle2019, Kosugi2022, Xie2024} under a Hamiltonian $H$. We will implement the evolution by constructing exact circuits in spectral space, exploiting the regular nature of diffusion on a uniform grid.

Simulating the discrete spectral operator $\exp(-Dk_j^2t)$ requires implementing the exponential of the squared wavenumber. We will consider Fourier space as the spectral space to impose periodic boundaries. Other boundary conditions follow later in the section. The squared wavenumbers for the QFT are defined and ordered by
\begin{equation}
    k_j^2 = \left(\frac{2\pi}{L}\right)^2
    \begin{cases}
        j^2 & \text{when } 0\leq j < \frac{N}{2}, \\
        (j-N)^2 & \text{when } \frac{N}{2}\leq j < N .
    \end{cases}
    \label{eq:squared_wavenumbers}
\end{equation}
Defining $\beta = Dt(2\pi/L)^2$ as the Fourier number $\text{Fo} = Dt/L^2$ scaled by $4\pi^2$ quantifying the number of diffusion time scales that have elapsed, the non-unitary evolution operator can be written as the piecewise expression
\begin{equation}
    e^{-Dk_j^2t} =
    \begin{cases}
        e^{-\beta j^2} & \text{when } 0\leq j < \frac{N}{2}, \\
        e^{-\beta(j-N)^2} & \text{when } \frac{N}{2}\leq j < N.
    \end{cases}
    \label{eq:diffusion_exp}
\end{equation}
Since the index $j$ has the binary expansion in Eq.~\eqref{eq:binary_expansion}, $j^2$ has the expansion
\begin{align}
    j^2 &= \left(\sum_{r=0}^{n-1}2^rq_r\right)^2 = \sum_{r=0}^{n-1}2^{2r}q_r + 2\sum_{s>r}^{n-1} 2^{r+s}q_rq_s.
\end{align}
Therefore, the non-unitary evolution for $j<N/2$ can be written as 
\begin{equation}
    e^{-\beta j^2} = \left[ \prod_{r=0}^{n-1} e^{-2^{2r}\beta q_r} \right] \left[ \prod_{s>r}^{n-1} e^{- \,2^{1+r+s}\beta q_r q_s} \right].
    \label{eq:exp_jsquared} 
\end{equation}
These products can be implemented in a quantum circuit using a unitary block encoding since all of the exponents are negative, ensuring that the magnitudes are less than one. However, when following the same procedure for the $j\geq N/2$ term in Eq.~\eqref{eq:diffusion_exp} with $(j-N)^2 = j^2-2Nj+N^2$, positive exponents arise from the change of sign in the cross term $-2jN$. These correspond to an amplification in spectral space, and cannot be directly implemented as a block encoding in a unitary circuit. To overcome this, we propose a method that exploits the property of the modes for $j \geq N/2$ requiring the same processing as $j<N/2$, but in reverse order and with a shift to $j+1$. The squared wavenumbers in Eq.~\eqref{eq:squared_wavenumbers} for $j\geq N/2$ can therefore be written as $(2\pi/L)^2 (j+1)^2$ for $j=\{N/2-1, N/2-2, \dots, 0\}$, where the index for $j\geq N/2$ is simply the reverse of the index for $j<N/2$. Consequently, this index has an entirely positive binary expansion of
\begin{align}
    (j+1)^2 &= \left(\sum_{r=0}^{n-1} 2^r q_r + 1 \right)^2 = \sum_{r=0}^{n-1} 2^{2r} q_r + 2 \sum_{r<s} 2^{r+s} q_r q_s + 2 \sum_{r=0}^{n-1} 2^r q_r + 1,
\end{align}
such that the exponents of $e^{-\beta(j+1)^2}$ are negative and can be block encoded in a unitary circuit. This exponential is given by the product
\begin{equation}
    e^{-\beta (j+1)^2} = 
    \left[ \prod_{r=0}^{n-1} e^{-2^{2r} \beta q_r} \right]
    \left[ \prod_{r < s} e^{-2^{1+r+s} \beta q_r q_s} \right]
    \left[ \prod_{r=0}^{n-1} e^{-2^{r+1} \beta q_r} \right]
    e^{-\beta}, 
    \label{eq:exp_jp1_squared}
\end{equation}
with the first two product terms being equivalent to the expression for $e^{-\beta j^2}$ in Eq.~\eqref{eq:exp_jsquared}.
\begin{figure}
    \centering
    \includegraphics[width=0.99\linewidth]{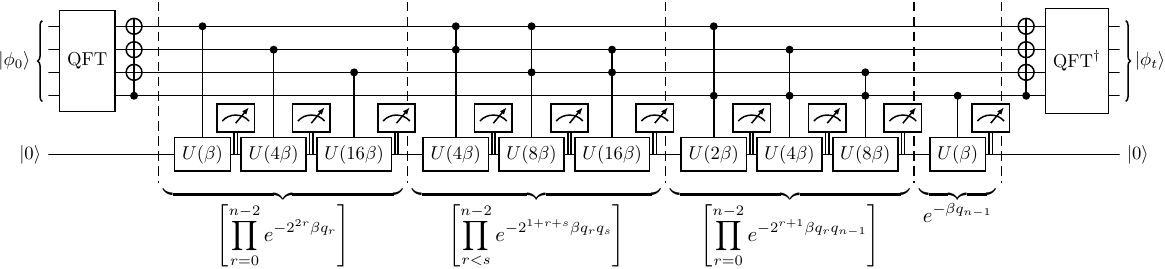}
    \caption{Quantum circuit for solving the one-dimensional heat equation with spectral accuracy on $N=16$ grid points with $O(\log^2N)$ gates. Periodic boundary conditions are used. The circuit implements the products in Eqs.~\eqref{eq:exp_jsquared} and \eqref{eq:exp_jp1_squared}. The unitary $U$ is defined in Eq.~\eqref{eq:U}, and the ancilla is measured to be $\ket{0}$ after each application. The outer CNOT gates flip the second half of the spectrum to exploit the mirroring of the squared wavenumbers, as discussed in the text.}
    \label{fig:diff_spectral}
\end{figure}
To implement the circuit, it is convenient to define the special case of the $R_Y$ gate
\begin{equation}
    U(\gamma) = R_Y\left[ 2 \arccos \left( e^{-\gamma} \right)\right] = \begin{bmatrix}
        e^{-\gamma} & -\sqrt{1-e^{-2\gamma}} \\
        \sqrt{1-e^{-2\gamma}}  & e^{-\gamma}
    \end{bmatrix},
    \label{eq:U}
\end{equation}
where $R_Y$ is the real orthogonal rotation
\begin{equation}
    R_Y(\theta) =
    \begin{bmatrix}
         \cos(\frac{\theta}{2}) & -\sin(\frac{\theta}{2}) \\  \sin(\frac{\theta}{2}) & \ \ \cos(\frac{\theta}{2})
    \end{bmatrix}.
    \label{eq:ry_gate}
\end{equation}
This is because the controlled version of this gate can implement the non-unitary damping $e^{-\gamma}$ when applied to an ancilla qubit in the $\ket{0}$ state, conditional on measuring the ancilla qubit to again be in the $\ket{0}$ state. A similar strategy was employed in a recent probabilistic imaginary time evolution algorithm\cite{Xie2024} where it is applied term-by-term after Pauli-string Trotterization. We use the technique between forward and inverse QFTs, thereby implementing the entire diffusive propagator in a single step. Figure \ref{fig:diff_spectral} shows the quantum circuit that implements the full operator using the decompositions given in Eqs.~\eqref{eq:exp_jsquared} and \eqref{eq:exp_jp1_squared}. The non-unitary product terms are implemented by applying a controlled rotation $U(\gamma)$ on an ancilla qubit in the $\ket{0}$ state, where $\gamma$ encodes the coefficients in Eqs.~\eqref{eq:exp_jsquared} and \eqref{eq:exp_jp1_squared}, and the control qubits are determined by $q_r$ and $q_s$. Postselecting on the ancilla measurement outcome $\ket{0}$ after each gate removes the unwanted entanglement for the next operation. Alternatively, a fresh ancilla qubit can be introduced for each operation with measurement performed at the end of the computation, but this would increase the qubit requirements quadratically.

Since the first two product terms are identical for both piecewise expressions in Eqs.~\eqref{eq:exp_jsquared} and \eqref{eq:exp_jp1_squared}, this operation can be implemented across both halves of the state by excluding the most-significant main register qubit from the operations. Then, the additional product terms for $j\geq N/2$ correspond to controlling operations by this qubit being in the $\ket{1}$ state. The technique to reverse the wavenumber indexing for $j\geq N/2$ is achieved by surrounding the entire transformation with CNOT gates, controlled by the most-significant main register qubit.

The diffusion operator can be implemented with $O(\log N)$ qubits and $O(\log^2 N)$ gates, where the spectral processing has the same gate complexity as the spectral transformation (QFT) circuit. In terms of the error where $N = \log(1/\epsilon)$, the circuit requires $O(\log^2 \log 1/\epsilon)$ gates. The gate complexity is independent of $t$ as this parameter only affects the angle of rotation in the $R_Y$ gates. Since the quantum circuit constructs the differential operators exactly in spectral space, the probability of success is $p(t) = \|\vec{\phi}(t)\|^2/\|\vec{\phi}(0)\|^2$, where $\vec{\phi}(t)$ is the unnormalized solution. For sufficiently large values of $\beta$, $\vec{\phi}(t)$ converges to $\text{mean}\{\vec{\phi}(0)\}$, thereby limiting the worst-case probability of success to $p_\infty = N|\text{mean}\{\vec{\phi}(0)\}|^2/\|\vec{\phi}(0)\|^2$.

\subsection{Implementation of Neumann and Dirichlet boundary conditions}
The methodology described in the previous sections is capable of simulating various computational boundary conditions due to the versatility of other discrete spectral transforms, such as the discrete cosine transform and the discrete sine transform. Both of these transformations are unitary and can be implemented by efficient quantum circuits, known as the quantum cosine transform (QCT) and quantum sine transform (QST).\cite{Klappenecker2001}

Homogeneous Neumann (zero-gradient) boundary conditions can be implemented by considering even symmetry around the computational boundaries. This can be implemented by substituting the QFT and QFT$^\dagger$ operations in Fig.~\ref{fig:diff_spectral} with the QCT and QCT$^\dagger$, respectively. For this, we selected the most common type-II transform,\cite{Strang1999} corresponding to $\phi_{-1}=\phi_0$ and $\phi_N = \phi_{N-1}$, where $\phi_0$ and $\phi_{N-1}$ correspond to the boundary nodes, and $\phi_{-1}$ and $\phi_{N}$ correspond to ghost nodes beyond the boundary. This QCT operator is defined by
\begin{equation}
    \text{QCT}_{kn} =
    \begin{cases}
        \displaystyle \sqrt{\frac{1}{N}} & \text{when } k = 0 \\
        \displaystyle \sqrt{\frac{2}{N}} \cos\left[ \frac{\pi}{N} \left( n + \frac{1}{2} \right) k \right] & \text{otherwise},
    \end{cases}
\end{equation}
for $k,n\in\{0, 1, \dots, N-1\}$. The QCT can be efficiently implemented, in essence, by applying a QFT of size $2N$ on a symmetric extension of the input.\cite{Klappenecker2001} Because the reflected signal is even, all of the sine modes vanish and only cosine modes remain. This can be implemented with the same $O(n\log[n/\epsilon])$ gate complexity as the QFT,\cite{Childs2021} but with one additional ancilla qubit to perform the unitary reflection.\cite{Klappenecker2001} The reflected domain is discarded by uncomputation after the operation.

Homogeneous Dirichlet (i.e.\ zero-valued) boundary conditions can be implemented by considering odd symmetry around the computational boundaries. The quantum sine transform (QST) of type-II corresponding to $\phi_{-1} = -\phi_{0}$ and $\phi_{N} = -\phi_{N-1}$ is defined as
\begin{equation}
    \text{QST}_{kn} = \sqrt{\frac{2}{N}} \sin\left[ \frac{\pi}{N} \left(n+1\right)\left(k+\frac{1}{2}\right) \right],
\end{equation}
for $k,n\in\{0, 1, \dots, N-1\}$. The aforementioned QCT algorithm implements the unitary $[\text{QCT}, 0; 0, -i\,\text{QST}]$\cite{Klappenecker2001} as block-diagonal. The initial state of the ancilla therefore determines the encoded condition, where $\ket{0}$ enacts the QCT and $\ket{1}$ enacts the QST after a phase correction. The ancilla remains in its initial unentangled state after the operation. While both operations are unitary, there are no efficient known circuits that can efficiently construct the transformation without an ancilla qubit, to the best of our knowledge. Inhomogeneous Dirichlet boundary conditions can also be encoded with the QST by evolving the fluctuation of the scalar about the steady state $\phi' = \phi-\overline{\phi}$, where the steady state scalar field $\overline{\phi}(x,y)=\phi_\text{off} + x\,\partial_x\overline{\phi} + y\,\partial_y\overline{\phi}$ varies under the constant linear gradients $\partial_x\overline{\phi}$ and $\partial_y\overline{\phi}$  and constant offset $\phi_\text{off}$.  The steady state $\bar{\phi}$ does not affect the time evolution as a constant offset and gradient vanish for a spatial second-order derivative. An advection-diffusion simulation of a shear flow with inhomogeneous boundaries in the $y$ direction is thus possible as $\vec{u}=[u(y),0]$ and the mean scalar $\overline{\phi}(y)$ is a function of $y$ only. 

When implementing Neumann or Dirichlet boundary conditions via the QCT or QST respectively, the wavenumbers are defined and ordered as $k_j = \pi j/L$ for Neumann conditions and $k_j = \pi(j+1)/L$ for Dirichlet conditions, where $j\in\{0, 1, \dots N-1\}$. The absence of negative wavenumbers simplifies the quantum circuit implementation, where diffusion can be implemented simply by the two product terms in Eq.~\eqref{eq:exp_jsquared}. For simulations involving various boundary conditions, the definition of $\beta$ is updated to $\beta_1=Dt(2\pi/L)^2$ for periodic boundary conditions, and $\beta_2 = Dt(\pi/L)^2$ for Neumann or Dirichlet boundary conditions, which is simply the Fourier number $\text{Fo} = Dt/L^2$ scaled by $4\pi^2$ and $\pi^2$ respectively.

\section{Spectral quantum advection-diffusion}

So far, quantum circuits for the individual simulation of advection and diffusion have been introduced. When both the advection and heat equations are diagonalizable by the same spectral operators, the Fourier-transformed operators commute and the individual implementations $\exp(-iuk_jt)\exp(-Dk_j^2t)\ket{\hat{\phi}_j}$ correspond to the combined dynamics of $\exp(-iuk_jt-Dk_j^2t)\ket{\hat{\phi}_j}$. However, this condition only holds for very simple one-dimensional flows. 

A more general algorithm for simulating scalar transport can be obtained by combining the advection equation and heat equation algorithms into an algorithm for the advection-diffusion equation by operator splitting. In the simplest case for an evolution time $t$ discretized by $N_t$ time steps, the Lie-Trotter product formula is
\begin{equation}
    e^{t(X+Y)} = \lim_{N_t\to \infty} \left(e^{tX/N_t} e^{tY/N_t}\right)^{N_t},
\end{equation}
where $X$ and $Y$ are the non-commuting right-hand-side operators for advection and diffusion, respectively. Defining the time step $\Delta t = t/N_t$, the error of the method can be estimated with the Baker-Campbell-Hausdorff expansion\cite{Magnus1954}
\begin{equation}
    \log\left(e^{X\Delta t}e^{Y\Delta t}\right) = X\Delta t + Y\Delta t + \frac{\Delta t^2}{2}[X,Y] + O(\Delta t^3),
\end{equation}
where the commutator $[X,Y] = XY-YX$. The local truncation error term is $O(\Delta t^2)$, resulting in first-order accuracy across the entire simulation time $t$. The error of the operator splitting can be estimated by analyzing the leading-order error term, $[X,Y]\Delta t^2/2$, although the assumption that the leading-order terms are dominant is not always valid.\cite{Childs2019}  When considering the two-dimensional advection-diffusion equation of a shear flow where $X = -u(y)\partial_x$ and $Y = D(\partial^2_x + \partial^2_y)$, we have
\begin{align}
    [X,Y]\phi &= -u(y)\partial_x\left(D[\partial_x^2\phi + \partial_y^2\phi]\right) - D\left(\partial_x^2[-u(y)\partial_x\phi] + \partial_y^2[-u(y)\partial_x\phi]\right) \nonumber \\
    &= D\left(2\,u'(y)\,\partial_{xy}\phi + u''(y)\,\partial_x\phi\right), \label{eq:commutator_error}
\end{align}
revealing that errors from the velocity shear $u'(y)$ grow with the mixed gradient $\partial_{xy}\phi$, as shear flows augment $\partial_y\phi$ at a rate that is dependent on $\partial_x\phi$. An additional error source arises from the product of the velocity curvature $u''$ with the streamwise gradient $\partial_x\phi$. As expected, the error vanishes for uniform velocity fields or when $\phi$ varies as a function of $y$ only. The accuracy can be improved to second-order using Strang splitting,
\begin{equation}
    e^{t(X+Y)} = \lim_{N_t\to \infty} \left(e^{tX/2N_t} e^{tY/N_t} e^{tX/2N_t}\right)^{N_t},
\end{equation}
and to arbitrarily higher orders of accuracy.\cite{Yoshida1990} However, orders of accuracy greater than two necessarily involve negative time steps,\cite{Blanes2005} presenting challenges for parabolic PDEs like the heat equation. This can be overcome with complex time steps with a positive real part,\cite{Blanes2010} although its quantum implementation falls outside the scope of this work. 
For second-order Strang splitting, $N_t = O(t^{3/2}\epsilon^{-1/2})$ since $\Delta t = O([\epsilon/t]^{1/2})$ and $N_t = t/\Delta t$, so solving the advection-diffusion equation requires $O(N_th\log^{h+1}N) = O(t^{3/2}\epsilon^{-1/2}h\log^{h+1} N)$ two-qubit gates, where the $h$ is the order of the polynomial velocity profile and its dependence was derived earlier for advection. Performing the exact time evolution for each operator improves over explicit time-marching methods that must adhere to the Courant-Friedrichs-Lewy condition. This allows the gate complexity to be logarithmic in $N$, compared to polynomial in $N$ for quantum explicit time-marching\cite{Over2025b} or classical methods.

\begin{figure}[t]
    \centering
    \includegraphics[width=0.99\textwidth]{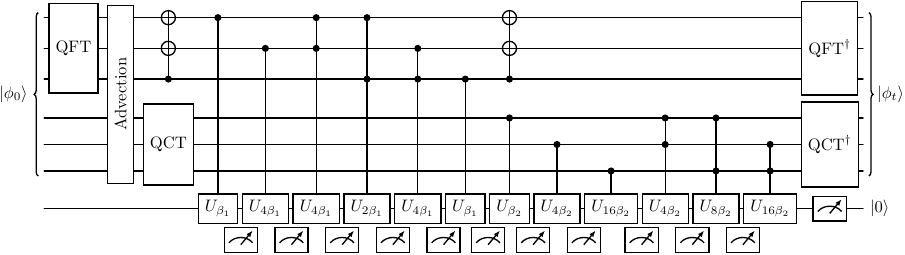}
    \caption{Quantum circuit for simulating a Trotter step of the advection-diffusion equation in a two-dimensional laminar shear flow using $n=3$ qubits per spatial dimension, where the top three qubits correspond to the $x$ direction and the next three qubits correspond to the $y$ direction. Ancilla qubit measurements of $\ket{0}$ are required for successful execution. The boundaries are periodic in $x$ and homogeneous Neumann in $y$, leading to $\beta_1=Dt(2\pi/L)^2$ and $\beta_2=Dt(\pi/L)^2$. Example advection block implementations are shown in Fig.~\ref{fig:advection_shear_circuits}. While the QCT is a unitary transformation, existing efficient circuits require one additional ancilla.}
    \label{fig:operator_splitting}
\end{figure}

The quantum circuit for solving the advection-diffusion equation in a laminar shear flow by Trotter splitting is shown in Fig.~\ref{fig:operator_splitting}. The domain is periodic in the $x$ direction, so the QFT is used as the spectral operator. At the boundaries with respect to $y$, the zero-scalar-flux condition $\partial \phi/\partial y=0$ is applied, so the QCT is used as the spectral operator for the $y$ register. The `Advection' block may represent any of the controlled phase sequences in Fig.~\ref{fig:advection_shear_circuits} to implement the desired velocity profile. The QCT has been depicted acting on $n$ qubits for simplicity and to emphasize its unitary nature, although its current most efficient gate decomposition requires an ancilla qubit for mirroring the state.\cite{Klappenecker2001} 

\section{Statevector simulations of advection-diffusion}

\subsection{Diffusive traveling pulse}

\begin{figure}[t]
    \centering
    \includegraphics[width=0.7\textwidth]{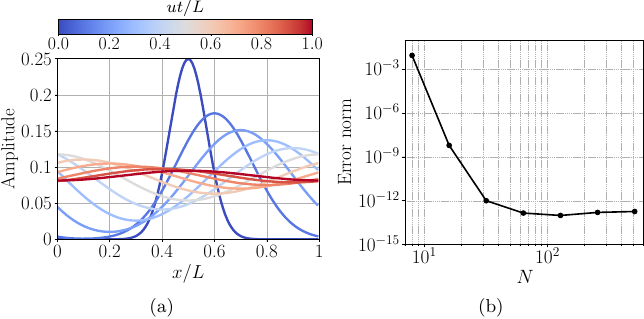}
    \caption{Statevector simulations of the one-dimensional advection-diffusion equation, showing (a) the evolution of the quantum amplitudes (and thus the solution) for $N=128$ grid points, and (b) the error norm versus grid size $N$. The error norm is obtained  with respect to the normalized solution of Eq.~\eqref{eq:analytical_solution}. The time evolution is done by a single time step.}
    \label{fig:diffusive_travelling_pulse}
\end{figure}

The baseline performance and accuracy of the algorithm will be validated in a one-dimensional statevector simulation with periodic boundary conditions. Here, the advection and diffusion operators commute, so full spectral accuracy can be achieved. The initial Gaussian function $\phi(0,x) = \exp(-100(x-0.5)^2)$ for $x\in[0,1)$ is encoded in the solution register
\begin{equation}
    \ket{\phi(0)} = \frac{1}{\sqrt{\sum_{j=0}^{N-1} \phi(0, j/N)^2}} \sum_{j=0}^{N-1} \phi(0,j/N) \ket{j},
\end{equation}
and is supplemented by one ancilla qubit for the overall initial state $\ket{0}\otimes \ket{\phi(0)}$. The fundamental solution to the heat equation is Gaussian, given by $\phi(t,x) = 1/\sqrt{4\pi Dt}\exp(-x^2/4Dt)$ in response to an initial Dirac delta condition $\phi(0,x) = \delta (x)$. Therefore, the diffusion algorithm itself can be used to prepare a Gaussian initial state by acting on a computational basis state, evolving under the conditions of $Dt=1/400$. However, the necessary boundary conditions restrict the range of usable states to cases where boundary values are negligible. The solution in a periodic computational domain of length $L$ is discretized using $N = \{8,16,32,\dots,512\}$ grid points. The simulation is evolved for $ut/L = 1$ for one complete pass of the domain with Fourier number $\text{Fo} = Dt/L^2 = 0.08$. This corresponds to a P\'eclet number $\text{Pe} = uL/D=12.5$ which is the ratio of the advective transport rate to the diffusive transport rate. The quantum circuit that implements the evolution is constructed by sequentially combining the Fourier-space operations shown in Figs.~\ref{fig:adv_spectral} and \ref{fig:diff_spectral} while being diagonalized under the same QFT operator. The analytical solution is
\begin{equation}
    \phi(t,x) = \int_{0}^{L} \left[ \frac{1}{\sqrt{4\pi D t}} \exp\left(-100 \left(\eta - 0.5\right)^2\right) \sum_{m=-\infty}^{\infty} \exp\left(-\frac{\left(x - u t - [\eta + m]\right)^2}{4 D t} \right) \right]d\eta,
    \label{eq:analytical_solution}
\end{equation}
resulting from the convolution of the Green function for the heat equation at constant advection with the initial condition of $\phi(0,x)$, and summing over integer shifts using the method of images to enforce periodic boundary conditions. The evolution of the quantum amplitudes in the solution register for the case where $N=128$ is shown in Fig.~\ref{fig:diffusive_travelling_pulse}a, and the greatest relative error norm at any stage of the simulation as a function of $N$ is shown in Fig.~\ref{fig:diffusive_travelling_pulse}b. After simulating the entire duration $ut/L=1$, the success probability is 25.1\% and corresponds to the theoretical $p(t) = \|\vec{\phi}(t)\|^2/\|\vec{\phi}(0)\|^2$ for this initial condition.

The error norm has been evaluated from $\|\ket{\phi(t)}-(\vec{\phi}(t)/\|\vec{\phi}(t)\| ) \|$ by comparing the postselected quantum statevector $\ket{\phi(t)}$ with the analytical solution vector $\vec{\phi}(t)$, which is then appropriately normalized. The error norm ranges between 0 for identical states and 2 for maximally different states. The simulation error rapidly converges to machine precision by approximately $N=32$ grid points due to the highly accurate spectral evaluation of the spatial derivatives. Furthermore, the simulation has considerable accuracy using just $N=8$ grid points, capturing the correct qualitative evolution with an error norm of approximately 0.009. 

\subsection{Scalar transport in laminar shear flows}

\begin{figure}[t!]
    \centering
    \includegraphics[width=0.99\textwidth]{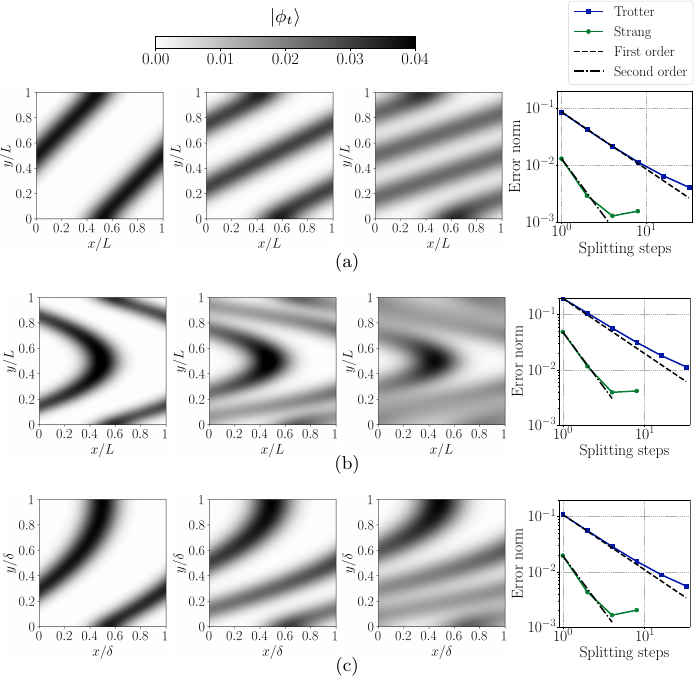}
    \caption{Quantum amplitudes computed with a second-order Strang splitting for the advection-diffusion equation at integer passes $Ut/L = 1,2,3$, with a splitting step $Ut/L = 0.5$. The error norms for both Trotter and Strang splitting methods are shown to the right at $Ut/L = 1$ for various time steppings. They are evaluated with respect to a reference solution, that is obtained via a tenth-order finite difference scheme. Results are presented for three flow configurations: (a) Couette flow, (b) channel flow, and (c) a third-order approximation of the Blasius boundary layer.}
    \label{fig:shear_contours}
\end{figure}

Statevector simulations of the quantum algorithm applied to the advection-diffusion equation in shear flows are presented in Fig.~\ref{fig:shear_contours}. The initial condition $\phi(0,x,y) = \exp(-100(x-0.5)^2)$ is identical to the one-dimensional simulations in the former example. The two-dimensional computational domain is square with side lengths $L$ and has been discretized with a uniform $64\times64$ grid, requiring 12 main register qubits and an ancilla qubit for a total of 13 qubits. The non-dimensional simulation time of $Ut/L=3$ corresponds to three passes at the location of maximum velocity $U$. The P\'eclet number is $\text{Pe} = UL/D = 500$, which has been substantially increased compared to the one-dimensional example, as shear flows enhance the formation of scalar gradients, significantly increasing the contribution of $(\partial_x^2 + \partial_y^2)\phi$ and therefore the rate of diffusion. The simulations are performed with Trotter splitting as shown in Fig.~\ref{fig:operator_splitting} or Strang splitting, which uses an additional advection block for second-order accuracy in time. The advection block was implemented using the three circuits in Fig.~\ref{fig:advection_shear_circuits} for Couette, plane Poiseuille, and Blasius boundary layer profiles. The simulations are periodic in the $x$ direction, and use Neumann boundary conditions in the $y$ direction with a zero gradient for all three cases. For a heat transfer problem, this corresponds to thermally insulated walls. The probability of simulating the entire duration $Ut/L=3$ is comparable to the one-dimensional pulse simulation due to the same initial condition, and depends on the rate of mixing induced by the shear flow. The probability of success is 33.3\% for the Couette flow, 30.3\% for the channel flow, and 35.7\% for the Blasius boundary layer, each corresponding to $p(t) = \|\vec{\phi}(t)\|^2/\|\vec{\phi}(0)\|^2$. Given the initial condition $\vec{\phi}(0)$, the final value of $\|\vec{\phi}(t)\|^2$ is determined by the Fourier number $\text{Fo} = Dt/L^2$ measuring how many characteristic diffusion time scales $L^2/D$ have elapsed, which is simply a scaled version of $\beta_1$ and $\beta_2$. This is independent of the number of advection time scales that have elapsed given by $Ut/L$, and therefore their ratio given by the P\'eclet number $\text{Pe} = UL/D$.

Since this configuration is non-trivial, no analytical reference solutions are available. Thus, the quantum solution is compared with a numerical solution obtained by a high-order finite difference method. This was evaluated on an identical $64\times 64$ grid using a tenth-order central finite difference scheme to evaluate the spatial derivatives. The Neumann boundary conditions were implemented using a ghost-cell method that corresponds to a reflection at half the grid spacing at $y=-\Delta y/2$ and $y=L+\Delta y/2$. For example, for the boundary node $\phi_0$ and first interior node $\phi_1$, the corresponding ghost nodes are $\phi_{-1}=\phi_{0}$, $\phi_{-2}=\phi_{1}$. This does not reduce the accuracy at the boundary, and matches the QCT implementation of type II. The dominant source of error in the shear flow configurations arises from the operator splitting because the one-dimensional simulations in Fig.~\ref{fig:diffusive_travelling_pulse} show that the errors rapidly converge to machine precision for commuting operators. The laminar shear flow simulations have been conducted for both Trotter (first-order) and Strang (second-order) operator splitting. The error norm is defined identically to the one-dimensional case and shown for a single pass of the domain together with the respective contours in Fig.~\ref{fig:shear_contours} with the number of steps. The smallest errors are obtained for the Couette flow and the largest errors are obtained for the channel flow, consistent with the value of the maximum velocity gradient for each flow and the theoretical expression in Eq.~\eqref{eq:commutator_error}. In all cases, the error norm scales with the theoretical orders of the Trotter and Strang splitting up to approximately $10^{-3}$ where the accuracy of the classical reference solution is reached. At this stage, a further decrease of  $\Delta t$ continues to reduce the error according to the order of accuracy up until the spectral method becomes the dominant error source, but this is not visible in the plot as the operator splitting solution has become more accurate than the reference finite difference solution.

\section{End-to-end hardware execution}
\begin{figure}
    \centering
    \includegraphics[width=0.99\linewidth]{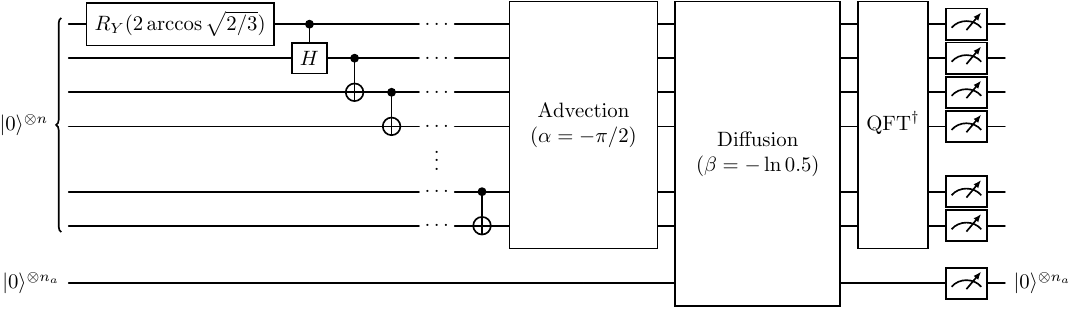}
    \caption{Quantum circuit and parameters for simulating the one-dimensional advection-diffusion equation with periodic boundaries on the real quantum devices. The initial state in the Fourier space, which is given by Eq.~\eqref{eq:initial_fourier_state}, is encoded prior to the advection block. The advection block is implemented as in Fig.~\ref{fig:adv_spectral}, and diffusion block is implemented as in Fig.~\ref{fig:diff_spectral}, but with an additional ancilla qubit for each controlled $R_Y$ gate. This meets the specific hardware requirements of measuring all qubits at the end of the computation.}
    \label{fig:hardware_circuit}
\end{figure}
\begin{figure}
    \centering
    \includegraphics[width=0.92\linewidth]{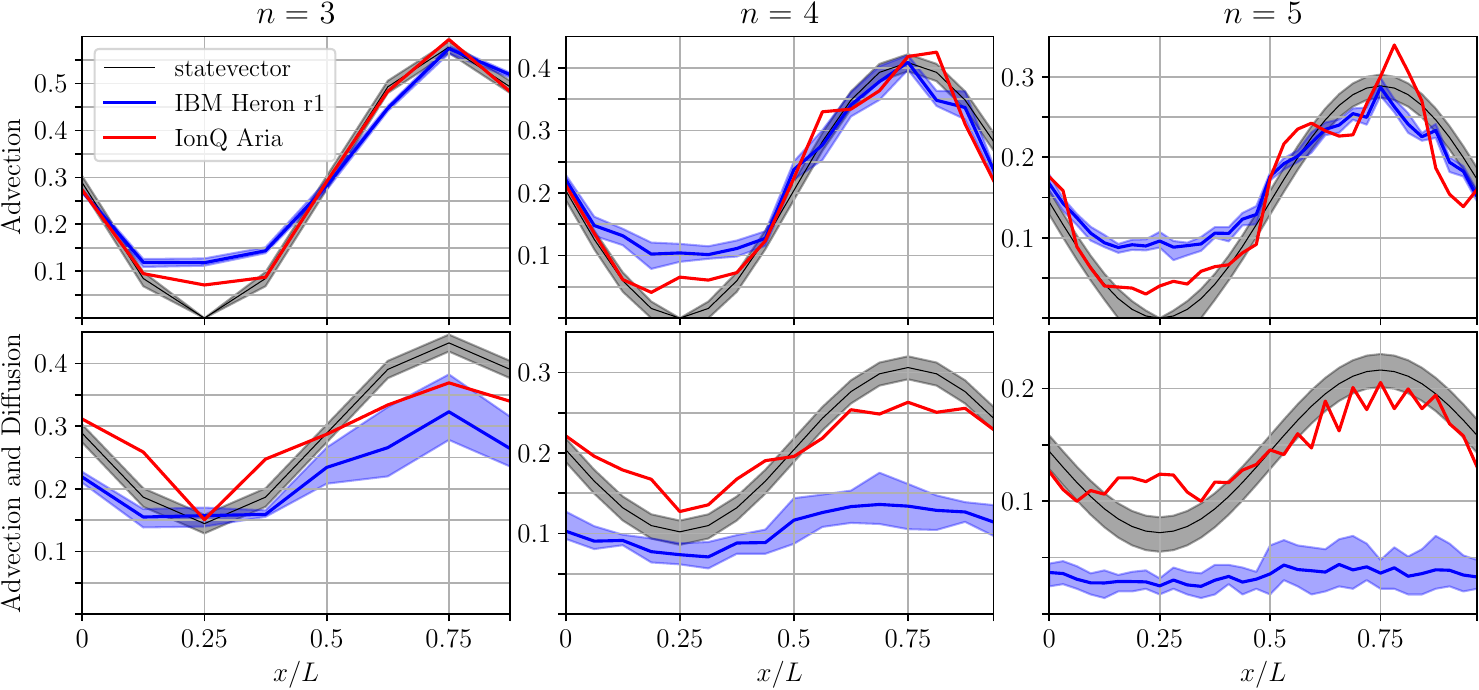}
    \caption{Quantum statevector reconstruction by means of $10^4$ shots (i.e., repeated runs of the simulation with measurement starting with identical initial conditions to determine the $n$-qubit quantum state) from a one-dimensional simulation on real quantum processors for pure advection (top row) and combined advection and diffusion (bottom row). The quantum simulations have been executed for $N=2^n$ grid points on $n=3,4,5$ qubits in the main register. The profile has been transported for $L/4U$ time units. The target solution (black) is compared against the solutions from a superconducting (IBM Heron, blue) and trapped-ion (IonQ Aria, red) platform. Solid lines are the reconstructed solutions. For the IBM Heron, the shaded area displays the range reaching from the minimum to the maximum deviation in an ensemble of 10 successive production runs. For the IonQ Aria, a single production run was possible only, such that the uncertainty range cannot be plotted. For the ideal statevector simulations, the shaded area corresponds to the range of a standard deviation of $\pm3\sigma$.}
    \label{fig:exe}
\end{figure}

Finally, we execute the algorithm on two different real quantum processors to evaluate the prospects on near-term quantum devices. We use the IBM Heron r1 (`ibm\_torino') with 133 nearest-neighbor-connected superconducting qubits and the IonQ Aria with 21 all-to-all trapped-ion qubits. 

The first step is state preparation. Since Fourier space simplifies many signals and particularly those composed of simple sines or cosines, the initial state was prepared directly in Fourier space. This removes an initial QFT and allows the preparation of an initial state proportional to $0.5(1+\cos x)$ directly as a sparse spectral representation. A one-dimensional periodic domain of length $L=2\pi$ is discretized by $N=8$, $16$ and $32$ grid points, requiring $n=3$, $4$ and $5$ qubits in the main register, respectively. The Fourier-transformed function has Fourier coefficients of $0.5$ at wavenumber $k=0$, and $0.25$ at $k=\pm 1$.  Figure~\ref{fig:hardware_circuit} shows the circuit to prepare this spectrum as well as the general workflow for the quantum simulations. In normalized form, this corresponds to the initial state
\begin{equation}
\ket{\hat{\phi}(0)} = \sqrt{\frac{2}{3}}\ket{0}^{\otimes n} + \sqrt{\frac{1}{6}} \left(\ket{0}^{\otimes (n-1)} \otimes \ket{1} + \ket{1}^{\otimes n}\right).
\label{eq:initial_fourier_state}
\end{equation}
The little-endian convention is used, where the least-significant qubit is at the top of circuit diagrams and last in the tensor product. This can be prepared by $O(\log N)$ two-qubit gates. The initial $R_Y$ single-qubit rotation gate in Fig.~\ref{fig:hardware_circuit} splits the unit amplitude to $\sqrt{2/3}$ on $\ket{0}^{\otimes n}$ and $\sqrt{1/3}$ on $\ket{0}^{\otimes (n-1)} \otimes \ket{1}$. By a controlled Hadamard gate $H$, we split the amplitude of $\sqrt{1/3}$ to $\sqrt{1/6}$ on $\ket{0}^{\otimes (n-1)}\otimes \ket{1}$ and $\ket{0}^{\otimes (n-2)}\otimes \ket{11}$. The remaining task is to incorporate $\ket{1}^{\otimes n}$ into the superposition by the chain of CNOT gates in Fig.~\ref{fig:hardware_circuit}  to correct the trailing zeros to ones.

The initial state is evolved by advection using the phase gates shown in Fig.~\ref{fig:adv_spectral} with $\alpha=-\pi/2$ corresponding to a leftwards quarter pass of the domain. At this stage, we applied the $\text{QFT}^\dagger$ and measured every main-register qubit to inspect the performance for pure advection before incorporating diffusion. The statevector is reconstructed from a production run with $M=10^4$ shots (or repeated simulation runs with measurement) for each case and shown in the first row of Fig.~\ref{fig:exe}. The noise-free statevector, the target solution of the problem, is illustrated as a black solid line for reference in each panel. We realized an ensemble of 10 production runs on the superconducting IBM Heron platform and a single run on the trapped-ion platform. The shaded areas show the minimum and maximum deviations over all runs. As the statevector provides the accurate measurement probability $p$ for each state, we interpret the corresponding minimum and maximum statistically. With the binomial standard deviation $\sigma=\sqrt{p(1-p)/M}$ after $M=10^4$ measurements or shots, the black shaded area corresponds to the amplitudes by $\sqrt{p\pm 3\sigma}$ in analogy to the empirical $3\sigma$-rule. This incorporates more than $99\%$ of all realizations under statistical noise. We confirmed this for all given statevectors by classical sampling for verification. For advection only, the qualitatively correct solution is obtained for all qubit architectures. The trapped-ion qubits of the IonQ Aria perform more accurately overall and better capture the extremities of the state. The superconducting qubits of the IBM Heron appear to add a diffusive effect by showing a less pronounced wave. While more accurate in magnitude, the IonQ Aria shows a significant wavering around the true solutions for the $n=4$ and $n=5$ cases, while the IBM Heron seems to have a smoother solution. 

In the next test, the diffusion block is implemented after advection. Each ancilla measurement is performed on a separate qubit, which can be implemented with the circuit in Fig.~\ref{fig:diff_spectral} using a new ancilla qubit per $R_Y$ gate. Mid-circuit measurements are available on some devices,\cite{Corcoles2021} but were unavailable for our experiment. On the one hand, additional qubits create a larger and thereby more challenging quantum system, while on the other hand, mid-circuit measurements expose the state to readout crosstalk, which arises from the strong interaction with the measured qubit while attempting to maintain isolation from the neighboring qubits.\cite{Gaebler2021} For diffusion, the parameter of $\beta=-\ln(0.5)$ was passed, corresponding to a decay of 50\% to the Fourier modes $j=1$ and $j=N-1$, resulting in a theoretical success probability of 75\%. We show the reconstructed statevector for the combined advection and diffusion, which corresponds to the ancilla qubits being in the state $\ket{0}^{\otimes n_a}$, in the second row of Fig.~\ref{fig:exe}. For $n=3$, $4$ and $5$, we need $n_a = 6$, $10$ and $15$ ancilla qubits, respectively. This results in a total qubit number of $9$, $14$ and $20$ for the three resolutions. The illustrated data is not normalized as the output on ancilla qubits can differ from $\ket{0}^{\otimes n_a}$. The results obtained for $n=3$ show similar qualitative results for the IBM and IonQ devices, but the IonQ device shows a greater success probability of 69\% compared to 40\% for the IBM device, which is closer to the theoretical 75\%. As the number of the qubits in the main register increases to $n=4$ and $n=5$, this trend continues, with the trapped-ion qubits of the IonQ Aria succeeding with 71\% and 70\% success probability, compared with 18\% and 3.6\% for the superconducting qubits. Note that for the $n=5$ result on the IonQ Aria, we manually addressed a coherent error which placed the solution not on state $\ket{0}^{\otimes n_a}$ but $\ket{1} \otimes\ket{0}^{\otimes( n_a-1)}$ on the ancilla qubits.

Overall, our experiments suggest that better performance is obtained for the trapped-ion architecture. The IBM Heron r1 executes the algorithm faster with a median two-qubit gate time of 84\,ns,\cite{Miessen2024} which compensates for the shorter phase coherence time $T_2\approx 100$ \textmu s,\cite{Abughanem2024} resulting a capacity for up to ${\sim}10^3$ coherent two-qubit gates. The IonQ Aria has a longer two-qubit gate time of 600\,\textmu s, but also a longer coherence time of $T_2 \approx 1$\,s, which also results in a capacity for up to ${\sim}10^3$ coherent two-qubit gates.\cite{IonQAria} This is sufficient for our calculations by up to one order of magnitude assuming all-to-all connectivity. The reported two-qubit gate fidelity of the IBM Heron r1 and the IonQ Aria quantum processing units is also comparable, at 99.5\% and 99.6\% respectively.\cite{Abughanem2024, IonQAria} Therefore, the increased performance of the IonQ Aria device arises from its all-to-all connectivity and its low readout error of 0.39\%.\cite{IonQAria} This is in contrast with the nearest neighbor connectivity of the IBM Heron r1 and its median readout error of 2\%.\cite{Miessen2024} Nearest neighbor connectivity substantially increases the required number of two-qubit gates, as the long-range gates require qubits to be routed through the register by a series of SWAP operations, each requiring three CNOT gates to implement. Figures \ref{fig:exe} and \ref{fig:adv_spectral} show that state preparation and advection respectively can be implemented on nearest-neighbor qubits, but the diffusion implementation in Fig. \ref{fig:diff_spectral} and the final $\text{QFT}^\dagger$ operation require long-range interactions, benefiting the all-to-all trapped-ion architecture.

When comparing between the two platforms, we also emphasize that we did not use a comprehensive circuit optimization. In both cases, advection can be realized by just two phase gates $P(-\pi/2)$ and $P(-\pi)$, as successive gates (e.g $P(-2\pi)$) are simply the identity operator. We kept the SWAP gates in the inverse QFT to have one full and regular QFT execution. Our initial condition given in Eq.~\eqref{eq:initial_fourier_state} is strongly localized in Fourier space, so the decay in Fourier modes is only relevant by the first and last ancilla rotation in Fig~\ref{fig:diff_spectral}. As the initial CNOT gates for periodic diffusion shift the initial amplitude of $\ket{1}^{\otimes n}$ to $\ket{1}\otimes\ket{0}^{\otimes (n-1)}$, there exists no state where two qubits are in state $\ket{1}$ simultaneously. Thus, all multi-controlled rotations are redundant for this initial condition. On the quantum processors, all ancilla rotations are fully implemented to replicate the effects of a more intricate initial state.

\section{Conclusions}

In this study, we have presented an efficient algorithm for simulating the advection-diffusion equation in laminar shear flows. This was achieved by diagonalizing individual dynamics with appropriate quantum spectral transforms, e.g.\ the QFT, then deriving explicit quantum circuits to simulate the resulting diagonal system $d\hat{\phi}_j/dt = zk_j^g\hat{\phi}_j$, where $z = -iu$ and $g=1$ for advection, and $z = -D$ and $g=2$ for diffusion. The systematic derivation of quantum circuits can equally be applied to any dynamics described by a negative real or imaginary $z$ and a positive integer $g$, such as for higher-order derivatives. The combined advection-diffusion dynamics were recovered by operator splitting, e.g. Trotter or Strang for first-order or second-order accuracy, respectively. While operator splitting divides the evolution into discrete time steps, it avoids the stability constraints of explicit time-marching methods as the splitting step $\Delta t$ remains independent of the grid size $\Delta x$. This has the crucial effect of decoupling the number of time steps and therefore the circuit depth from the number of points on the grid, thus retaining the exponential improvement over classical methods. The algorithm was demonstrated by statevector simulation for two-dimensional shear flows and on superconducting and trapped-ion quantum processors for one-dimensional flows.

Spectral methods efficiently diagonalize the differential operators, made possible by the regularity of the problem under a constant transport coefficient (i.e.\ $u$ or $D$) and a uniform Cartesian mesh. Pure advection is a unitary process and, therefore, amenable to quantum computation, but it typically occurs under a turbulent and therefore non-constant velocity field, preventing diagonalization. Diffusion imposes the opposite challenge, as it is not unitary, but typically occurs under a constant diffusivity $D$, so it can be efficiently diagonalized. This fundamental difference in physical behavior is likely to reward evolving by individual dynamics with favorable approaches before being combined with operator splitting. We exploit the property of two-dimensional laminar shear flows of the form $u(y)$ that they can be considered as an ensemble of one-dimensional flows, and therefore can be diagonalized by applying the spectral transform in one dimension only. More intricate, turbulent velocity fields require replacing the advection implementation with more versatile but less efficient approaches.\cite{Brearley2024} While some irregularities can be studied, for instance non-uniform grids via Chebyshev polynomials, a compromise in efficiency can be expected.\cite{Bengoechea2024}

The implementation of the diffusion operator requires a measurement of the single ancilla qubit after each gate, resulting in $O(\log^2 N)$ total measurements. The success probability is independent of the number of measurements, as given by $\|\vec{\phi}(t)\|^2 / \|\vec{\phi}(0)\|^2$ of the unnormalized solution $\vec{\phi}(t)$. This requirement for frequent mid-circuit measurement may introduce readout crosstalk errors on real hardware.\cite{Gaebler2021} Adding more than one ancilla qubit allows the parallelization of many of the controlled rotations to reduce the circuit depth. If the amplitude loss is known precisely, amplitude amplification\cite{Brassard2002} can avoid intermediate measurements by amplifying the target output to precisely 100\%. Although this can release ancilla qubits for further use, the additional cost of deeper circuits and the appearance of amplitude loss under analytical and real conditions need to be further discussed for practical advantage.

Operator splitting can be applied to other dynamical systems evolving under different differential operators, such as non-linear systems with source terms, but it remains to be seen whether relevant dynamics in fluid mechanics can be realized without a significant amplitude loss. The conventional mitigation by amplitude amplification may become problematic,\cite{Zecchi2025} which leads to the important question of whether the prevalent non-unitarity of fluid flows is amenable to simulation on quantum computers. This work uses the desirable principle of evolving without requiring intermediate measurements and re-initializations of the solution register.\cite{Lu2024} As the solution cannot be visually inspected, we rely on the simulation to proceed without unexpected failures. The accuracy and efficiency of spectral methods may improve this aspect, but artifacts such as the Gibbs phenomenon may amplify imperfections in the state. The mildly dissipative nature of upwind finite difference schemes, while much less accurate on smooth data, make them more robust to noise and discontinuities in the solution.\cite{Brearley2024}

For any processing scheme, the overall strategy still assumes that the input and output data can be efficiently accessed, which remains a significant obstacle to achieving a practical quantum advantage.\cite{Succi2023} Initial states that are a finite-length Fourier series can be efficiently encoded by reconstructing the Fourier spectrum and applying the IQFT, as demonstrated in the previous section. Considering such restrictions, we expect that quantum computational fluid dynamics is likely to mature not as a wholesale replacement for classical solvers, but as a specialized tool that is most powerful when regularity and reducibility of input, processing, and output coincide.

\vspace{2em}
\section*{Acknowledgments}
P.B. and S.L. are supported by UK EPSRC grant EP/W032643/1, and the UK National Quantum Computing Centre [NQCC200921]. P.P. was supported by the project no.\ P2018-02-001 "DeepTurb -- Deep Learning in and of Turbulence" of the Carl Zeiss Foundation. The work of P.P. and J.S. is also supported by the European Union (ERC, MesoComp, 101052786). Views and opinions expressed are however those of the authors only and do not necessarily reflect those of the European Union or the European Research Council. Neither the European Union nor the granting authority can be held responsible for them.

\section*{Author contributions statement}
P.P. and P.B. designed the research and conducted the analysis. All authors discussed the results. All authors wrote and reviewed the manuscript.

\section*{Data availability}
Data and source files are available from the corresponding author upon reasonable request.

\section*{Competing Interests}
The authors report no conflict of interest.

\bibliography{references.bib}

@article{Abughanem2024,
  title={IBM quantum computers: evolution, performance, and future directions},
  author={AbuGhanem, Muhammad},
  journal={arXiv preprint arXiv:2410.00916},
  year={2024}
}

@article{An2023,
    title={Linear combination of {Hamiltonian} simulation for nonunitary dynamics with optimal state preparation cost},
    author={An, Dong and Liu, Jin-Peng and Lin, Lin},
    journal={Physical Review Letters},
    volume={131},
    number={15},
    pages={150603},
    year={2023},
    publisher={APS},
    url={https://doi.org/10.1103/PhysRevLett.131.150603}
}

@article{Arute2019,
    title={Quantum supremacy using a programmable superconducting processor},
    author={Arute, Frank and Arya, Kunal and Babbush, Ryan and Bacon, Dave and Bardin, Joseph C and Barends, Rami and Biswas, Rupak and Boixo, Sergio and Brandao, Fernando GSL and Buell, David A and others},
    journal={Nature},
    volume={574},
    number={7779},
    pages={505--510},
    year={2019},
    publisher={Nature Publishing Group},
    url={https://doi.org/10.1038/s41586-019-1666-5}
}

@article{Au-Yeung2024,
    title={Quantum algorithms for scientific computing},
    author={Au-Yeung, Rhonda and Camino, Bruno and Rathore, Omer and Kendon, Viv},
    journal={Rep. Prog. Phys.},
    volume={87},
    number={11},
    pages={116001},
    year={2024},
    publisher={IOP Publishing},
    url={https://doi.org/10.1088/1361-6633/ad85f0}
}

@article{Au-Yeung2025,
    title={Quantum smoothed particle hydrodynamics algorithm inspired by quantum walks},
    author={Au-Yeung, R and Kendon, VM and Lind, SJ},
    journal={Physics of Fluids},
    volume={37},
    number={5},
    year={2025},
    publisher={AIP Publishing}
}

@article{Barenco1995,
    title={Elementary gates for quantum computation},
    author={Barenco, Adriano and Bennett, Charles H and Cleve, Richard and DiVincenzo, David P and Margolus, Norman and Shor, Peter and Sleator, Tycho and Smolin, John A and Weinfurter, Harald},
    journal={Phys. Rev. A},
    volume={52},
    number={5},
    pages={3457},
    year={1995},
    publisher={APS}
}

@article{Benioff1980,
    title={{The computer as a physical system: A microscopic quantum mechanical {Hamiltonian} model of computers as represented by Turing machines}},
    author={Benioff, Paul},
    journal={J. Stat. Phys.},
    volume={22},
    pages={563--591},
    year={1980},
    publisher={Springer},
    url={https://doi.org/10.1007/BF01011339}
}

@article{Bengoechea2024,
    title={Towards Variational Quantum Algorithms for generalized linear and nonlinear transport phenomena},
    author={Bengoechea, Sergio and Over, Paul and Jaksch, Dieter and Rung, Thomas},
    journal={arXiv preprint arXiv:2411.14931},
    year={2024}
}

@article{Berry2024,
    title={Quantum algorithm for time-dependent differential equations using {Dyson} series},
    author={Berry, Dominic W and Costa, Pedro CS},
    journal={Quantum},
    volume={8},
    pages={1369},
    year={2024},
    publisher={Verein zur F{\"o}rderung des Open Access Publizierens in den Quantenwissenschaften},
    url={https://doi.org/10.22331/q-2024-06-13-1369}
}

@article{Bharadwaj2023,
    title={Hybrid quantum algorithms for flow problems},
    author={Bharadwaj, Sachin S. and Sreenivasan, Katepalli R.},
    journal={Proceedings of the National Academy of Sciences of the USA},
    volume={120},
    number={49},
    pages={e2311014120},
    year={2023}
}

@article{Bharadwaj2025,
    title={Compact quantum algorithms for time-dependent differential equations},
    author={S. S. Bharadwaj and K. R. Sreenivasan},
    journal={Phys. Rev. Res.},
    volume={7},
    pages={023262},
    year={2025}
}

@article{Blanes2005,
    title={On the necessity of negative coefficients for operator splitting schemes of order higher than two},
    author={Blanes, Sergio and Casas, Fernando},
    journal={Applied Numerical Mathematics},
    volume={54},
    number={1},
    pages={23--37},
    year={2005},
    publisher={Elsevier}
}

@article{Blanes2010,
    title={Splitting methods with complex coefficients},
    author={Blanes, Sergio and Casas, Fernando and Murua, Ander},
    journal={SeMA Journal},
    volume={50},
    number={1},
    pages={47--60},
    year={2010},
    publisher={Springer}
}

@incollection{Brassard2002,
    author       = {Brassard, Gilles and H{\o}yer, Peter and Mosca, Michele and Tapp, Alain},
    title        = {Quantum amplitude amplification and estimation},
    booktitle    = {Quantum Computation and Information},
    series       = {Contemporary Mathematics},
    volume       = {305},
    pages        = {53--74},
    publisher    = {American Mathematical Society},
    address      = {Providence, RI},
    year         = {2002},
    isbn         = {978-0-8218-2140-4},
    mrnumber     = {1947332},
    url          = {https://doi.org/10.1090/conm/305/05215}
}

@article{Brearley2024,
    title={{Quantum algorithm for solving the advection equation using {Hamiltonian} simulation}},
    author={Brearley, Peter and Laizet, Sylvain},
    journal={Physical Review A},
    volume={110},
    number={1},
    pages={012430},
    year={2024},
    publisher={APS},
    url={https://doi.org/10.1103/PhysRevA.110.012430}
}

@book{Bruus2007,
    author = {H. Bruus},
    title = {Theoretical Microfluidics},
    publisher = {Oxford University Press},
    year = {2007},
    address = {Oxford, UK}
}

@article{Budinski2021,
    title={Quantum algorithm for the advection-diffusion equation simulated with the lattice {Boltzmann} method},
    author={Budinski, Ljubomir},
    journal={Quantum Information Processing},
    volume={20},
    number={2},
    pages={57},
    year={2021},
    publisher={Springer},
    url={https://doi.org/10.1007/s11128-021-02996-3}
}

@article{Childs2019,
    title={Theory of Trotter error with commutator scaling},
    author={Childs, Andrew M and Su, Yuan and Tran, Minh C and Wiebe, Nathan and Zhu, Shuchen},
    journal={Physical Review X},
    volume={11},
    number={1},
    pages={011020},
    year={2021},
    publisher={APS}
}

@article{Childs2021,
    title={High-precision quantum algorithms for partial differential equations},
    author={Childs, Andrew M and Liu, Jin-Peng and Ostrander, Aaron},
    journal={Quantum},
    volume={5},
    pages={574},
    year={2021},
    publisher={Verein zur F{\"o}rderung des Open Access Publizierens in den Quantenwissenschaften}
}

@inproceedings{Cleve2000,
    title={Fast parallel circuits for the quantum Fourier transform},
    author={Cleve, Richard and Watrous, John},
    booktitle={Proceedings 41st Annual Symposium on Foundations of Computer Science},
    pages={526--536},
    year={2000},
    organization={IEEE}
}

@article{Corcoles2021,
    title={Exploiting dynamic quantum circuits in a quantum algorithm with superconducting qubits},
    author={C{\'o}rcoles, Antonio D and Takita, Maika and Inoue, Ken and Lekuch, Scott and Minev, Zlatko K and Chow, Jerry M and Gambetta, Jay M},
    journal={Physical Review Letters},
    volume={127},
    number={10},
    pages={100501},
    year={2021},
    publisher={APS}
}

@article{Demirdjian2022,
    title={Variational quantum solutions to the advection-diffusion equation for applications in fluid dynamics},
    author={Demirdjian, Reuben and Gunlycke, Daniel and Reynolds, Carolyn A and Doyle, James D and Tafur, Sergio},
    journal={Quantum Information Processing},
    volume={21},
    number={9},
    pages={322},
    year={2022},
    publisher={Springer},
    url={https://doi.org/10.1007/s11128-022-03667-7}
}

@article{Elrod1979,
    title={{A general theory for laminar lubrication with Reynolds roughness}},
    author={Elrod, Harold G.},
    journal={Journal of Lubrication Technology},
    volume={101},
    number={1},
    pages={8--14},
    year={1979},
    publisher={ASME}
}

@article{Fang2023,
    title={Time-marching based quantum solvers for time-dependent linear differential equations},
    author={Fang, Di and Lin, Lin and Tong, Yu},
    journal={Quantum},
    volume={7},
    pages={955},
    year={2023},
    publisher={Verein zur F{\"o}rderung des Open Access Publizierens in den Quantenwissenschaften},
    url={https://doi.org/10.22331/q-2023-03-20-955}
}

@article{Feynman1982,
    title={Simulating Physics with Computers},
    author={Feynman, Richard P},
    journal={International Journal of Theoretical Physics},
    volume={21},
    number={6/7},
    year={1982},
    url={https://doi.org/10.1007/BF02650179}
}

@article{Gaebler2021,
    title={Suppression of midcircuit measurement crosstalk errors with micromotion},
    author={Gaebler, John P and Baldwin, Charles H and Moses, Steven A and Dreiling, Joan M and Figgatt, Caroline and Foss-Feig, Michael and Hayes, David and Pino, Juan M},
    journal={Physical Review A},
    volume={104},
    number={6},
    pages={062440},
    year={2021},
    publisher={APS}
}

@inproceedings{Gilyen2019,
    title={Quantum singular value transformation and beyond: exponential improvements for quantum matrix arithmetics},
    author={Gily{\'e}n, Andr{\'a}s and Su, Yuan and Low, Guang Hao and Wiebe, Nathan},
    booktitle={Proceedings of the 51st Annual ACM SIGACT Symposium on Theory of Computing},
    pages={193--204},
    year={2019},
    url = {https://doi.org/10.1145/3313276.3316366}
}

@article{Harrow2009,
    title={Quantum algorithm for linear systems of equations},
    author={Harrow, Aram W and Hassidim, Avinatan and Lloyd, Seth},
    journal={Phys. Rev. Lett.},
    volume={103},
    number={15},
    pages={150502},
    year={2009},
    publisher={APS},
    url={https://doi.org/10.1103/PhysRevLett.103.150502}
}

@article{Ingelmann2024,
    title={Two quantum algorithms for solving the one-dimensional advection-diffusion equation},
    author={Ingelmann, Julia and Bharadwaj, Sachin S and Pfeffer, Philipp and Sreenivasan, Katepalli R and Schumacher, J{\"o}rg},
    journal={Comput. Fluids},
    volume={281},
    pages={106369},
    year={2024},
    publisher={Elsevier},
    url={https://doi.org/10.1016/j.compfluid.2024.106369}
}

@misc{IonQAria,
    title        = {{IonQ Aria: Practical Performance}},
    howpublished = {\url{https://ionq.com/resources/ionq-aria-practical-performance}},
    note         = {Last updated: January 8, 2025. Accessed: 2025-10-14}
}

@article{Jaksch2023,
    author = {Jaksch, Dieter and Givi, Peyman and Daley, Andrew J. and Rung, Thomas},
    title = {Variational Quantum Algorithms for Computational Fluid Dynamics},
    journal = {AIAA Journal},
    volume = {61},
    number = {5},
    pages = {1885-1894},
    year = {2023},
    URL = {https://doi.org/10.2514/1.J062426}
}

@article{Jin2023,
    title={Quantum simulation of partial differential equations: Applications and detailed analysis},
    author={Jin, Shi and Liu, Nana and Yu, Yue},
    journal={Physical Review A},
    volume={108},
    number={3},
    pages={032603},
    year={2023},
    publisher={APS},
    url={https://doi.org/10.1103/PhysRevA.108.032603}
}

@inproceedings{Klappenecker2001,
    title={Discrete cosine transforms on quantum computers},
    author={Klappenecker, Andreas and Rotteler, Martin},
    booktitle={ISPA 2001. Proceedings of the 2nd International Symposium on Image and Signal Processing and Analysis. In conjunction with 23rd International Conference on Information Technology Interfaces (IEEE Cat.},
    pages={464--468},
    year={2001},
    organization={IEEE}
}

@article{Koecher2025,
    title={{Numerical solution of nonlinear Schrödinger equation by a hybrid pseudospectral-variational quantum algorithm}}, 
    author={N. Köcher and H. Rose and S. S. Bharadwaj and J. Schumacher and S. Schumacher},
    journal={Sci. Rep.},
    year={2025},
    volume={15},
    pages={23478}
}

@article{Kosugi2022,
    title={Imaginary-time evolution using forward and backward real-time evolution with a single ancilla: First-quantized eigensolver algorithm for quantum chemistry},
    author={Kosugi, Taichi and Nishiya, Yusuke and Nishi, Hirofumi and Matsushita, Yu-ichiro},
    journal={Physical Review Research},
    volume={4},
    number={3},
    pages={033121},
    year={2022},
    publisher={APS}
}

@article{Linden2022,
    title={Quantum vs. classical algorithms for solving the heat equation},
    author={Linden, Noah and Montanaro, Ashley and Shao, Changpeng},
    journal={Communications in Mathematical Physics},
    volume={395},
    number={2},
    pages={601--641},
    year={2022},
    publisher={Springer},
    url={https://doi.org/10.1007/s00220-022-04442-6}
}

@article{Low2019,
    title={{Hamiltonian} simulation by qubitization},
    author={Low, Guang Hao and Chuang, Isaac L},
    journal={Quantum},
    volume={3},
    pages={163},
    year={2019},
    publisher={Verein zur F{\"o}rderung des Open Access Publizierens in den Quantenwissenschaften}
}

@article{Lu2024,
    title={Quantum computing of reacting flows via {Hamiltonian} simulation},
    author={Lu, Zhen and Yang, Yue},
    journal={Proceedings of the Combustion Institute},
    volume={40},
    number={1-4},
    pages={105440},
    year={2024},
    publisher={Elsevier}
}

@article{Lubasch2025,
    title={{Quantum circuits for partial differential equations in Fourier space}},
    author={M. Lubasch and Y. Kikuchi and L. Wright and C. {Mc Keever}},
    journal={arXiv preprint arXiv:2505.16895},
    year={2025}
}

@article{Magnus1954,
    title={On the exponential solution of differential equations for a linear operator},
    author={Magnus, Wilhelm},
    journal={Commun. Pure Appl. Math.},
    volume={7},
    number={4},
    pages={649--673},
    year={1954},
    publisher={Wiley Online Library}
}

@article{McArdle2019,
    title={Variational ansatz-based quantum simulation of imaginary time evolution},
    author={McArdle, Sam and Jones, Tyson and Endo, Suguru and Li, Ying and Benjamin, Simon C and Yuan, Xiao},
    journal={npj Quantum Information},
    volume={5},
    number={1},
    pages={75},
    year={2019},
    publisher={Nature Publishing Group UK London}
}

@article{Meng2023,
    title={{Quantum computing of fluid dynamics using the hydrodynamic Schrödinger equation}}, 
    author={Z. Meng and Y. Yang},
    journal={Phys. Rev. Res.},
    year={2023},
    volume={5},
    pages={033182}
}

@article{Miessen2024,
  title={Benchmarking digital quantum simulations above hundreds of qubits using quantum critical dynamics},
  author={Miessen, Alexander and Egger, Daniel J and Tavernelli, Ivano and Mazzola, Guglielmo},
  journal={PRX Quantum},
  volume={5},
  number={4},
  pages={040320},
  year={2024},
  publisher={APS}
}

@article{Motta2020,
    title={Determining eigenstates and thermal states on a quantum computer using quantum imaginary time evolution},
    author={Motta, Mario and Sun, Chong and Tan, Adrian TK and O’Rourke, Matthew J and Ye, Erika and Minnich, Austin J and Brandao, Fernando GSL and Chan, Garnet Kin-Lic},
    journal={Nature Physics},
    volume={16},
    number={2},
    pages={205--210},
    year={2020},
    publisher={Nature Publishing Group UK London}
}

@article{Over2025a,
    title={Boundary treatment for variational quantum simulations of partial differential equations on quantum computers},
    author={P. Over and S. Bengoechea and T. Rung and F. Clerici and L. Scandurra and E. {de Villiers} and D. Jaksch},
    journal={Comput. Fluids},
    volume={288},
    pages={106508},
    year={2025}
}

@article{Over2025b,
    title={Quantum algorithm for the advection-diffusion equation by direct block encoding of the time-marching operator},
    author={Over, Paul and Bengoechea, Sergio and Brearley, Peter and Laizet, Sylvain and Rung, Thomas},
    journal={Physical Review A},
    volume={112},
    number={1},
    pages={L010401},
    year={2025},
    publisher={APS}
}

@article{Sato2024,
    title={{Hamiltonian} simulation for hyperbolic partial differential equations by scalable quantum circuits},
    author={Sato, Yuki and Kondo, Ruho and Hamamura, Ikko and Onodera, Tamiya and Yamamoto, Naoki},
    journal={Physical Review Research},
    volume={6},
    number={3},
    pages={033246},
    year={2024},
    publisher={APS},
    url={https://doi.org/10.1103/PhysRevResearch.6.033246}
}

@book{Schlichting2016,
    author = {H. Schlichting and K. Gersten},
    title = {Boundary-Layer Theory},
    publisher = {Springer},
    year = {2016},
    address = {Berlin}
}

@inproceedings{Shor1994,
    title={Algorithms for quantum computation: discrete logarithms and factoring},
    author={Shor, Peter W},
    booktitle={Proceedings 35th Annual Symposium on Foundations of Computer Science},
    pages={124--134},
    year={1994},
    organization={IEEE},
    url={https://doi.org/10.1109/SFCS.1994.365700}
}

@article{Strang1999,
    author    = {Gilbert Strang},
    title     = {The Discrete Cosine Transform},
    journal   = {SIAM Review},
    volume    = {41},
    number    = {1},
    pages     = {135--147},
    year      = {1999},
    url       = {https://math.mit.edu/~gs/papers/dct.pdf}
}

@article{Succi2023,
    title={Quantum computing for fluids: Where do we stand?},
    author={Succi, Sauro and Itani, Wael and Sreenivasan, Katepalli and Steijl, Ren{\'e}},
    journal={Europhysics Letters},
    volume={144},
    number={1},
    pages={10001},
    year={2023},
    publisher={IOP Publishing}
}

@article{Villermaux2019,
	title={Mixing versus stirring},
	author={Villermaux, E.},
	journal={Annual Reviews of Fluid Mechanics},
	volume={51},
	pages={245--273},
	year={2019}
}

@article{Wright2024,
    title={Noisy intermediate-scale quantum simulation of the one-dimensional wave equation},
    author={L. Wright and C. {Mc Keever} and J. T. First and R. Johnston and J. Tillay and S. Chaney and M. Rosenkranz and M. Lubasch},
    journal={Phys. Rev. Res.},
    volume={6},
    pages={043169},
    year={2024}
}

@article{Xie2024,
    title={Probabilistic imaginary-time evolution algorithm based on nonunitary quantum circuits},
    author={Xie, Hao-Nan and Wei, Shi-Jie and Yang, Fan and Wang, Zheng-An and Chen, Chi-Tong and Fan, Heng and Long, Gui-Lu},
    journal={Physical Review A},
    volume={109},
    number={5},
    pages={052414},
    year={2024},
    publisher={APS}
}

@article{Yoshida1990,
    title={Construction of higher order symplectic integrators},
    author={Yoshida, Haruo},
    journal={Phys. Lett. A},
    volume={150},
    number={5-7},
    pages={262--268},
    year={1990},
    publisher={Elsevier}
}

@article{Zecchi2025,
    title={Improved amplitude amplification strategies for the quantum simulation of classical transport problems},
    author={Zecchi, Alessandro Andrea and Sanavio, Claudio and Perotto, Simona and Succi, Sauro},
    journal={Quantum Science and Technology},
    volume={10},
    number={3},
    pages={035039},
    year={2025},
    publisher={IOP Publishing}
}

\newpage
\appendix
\counterwithin{figure}{section}
\counterwithin{equation}{section}
\refstepcounter{section} 

\end{document}